\documentclass[11pt,twoside]{article}

\usepackage{asp2006}
\usepackage{epsf}
\usepackage{psfig}
\usepackage{lscape}
\usepackage{graphicx}
\usepackage{epstopdf}
\usepackage{subfigure}
\begin{document}
\title{The Hot and Cold Outflows of NGC 3079}   
\author{N. Shafi$^{1}$,  R. Morganti$^{2}$, T. Oosterloo$^{2}$, R. Booth$^{1}$}   
\affil{$^{1}$HartRAO and University of the Witwatersrand, South Africa $^{2}$ASTRON, The Netherlands}    

\begin{abstract} 
Very deep  neutral hydrogen (HI) observations of the edge-on spiral galaxy NGC 3079 with the Westerbork Synthesis Radio Telescope (WSRT) are presented. The galaxy has been studied extensively in 
different wavelengths and known for its several unique  and complex features. However, the new data  still revealed several new features and show that the galaxy is even more complicated and interesting 
than previously thought. In the new data a large stream of gas, encircling the entire galaxy, was discovered, while the data also show, for the first time, that not only hot gas is blown into space by the 
starburst/AGN, but also large amounts of cold gas, despite the high energies involved in the outflow.  
\end{abstract}

\section{NGC 3079}   
NGC 3079 is one of the nearby ($\sim16$ Mpc) edge-on spiral galaxies that have been studied extensively in different 
wavelengths. It is classified as Seyfert 2 \citep{Ford1986} and shows evidence of starburst activity \citep[e.g;][]{Veilleux1994}. The galaxy is known for its several unique features, including an ionized gas outflow as a "super-bubble" \citep{Cecil2001}. The outflows seem to originate from the nucleus and reach velocities up to $\sim 1500$ km s$^{-1}$ \citep{Cecil2001}. However the exact origin of this "super-bubble" feature is still ambiguous due to the difficulty to distinguish the role of the starburst from the AGN.  Chandra X-ray observations also show a clear correspondence of X-ray filaments with H$\alpha$ filaments extended $\sim 65$ pc from the nucleus \citep{Cecil2002}.
A nuclear jet outflow has also been observed in the radio continuum as two, kilo parsec-scale, radio lobes extending on both sides of the major axis of the galaxy \citep{Duric1988, Baan1995}. These radio lobes are believed to be associated with the nuclear activity of the galaxy.\\
It is  interesting to note that NGC 3079 is found in a small group with two known small companion galaxies: NGC 3073  and MCG-917-9. The optical appearance of both companions do not show any disturbance, however, in radio NGC 3073 seems to be affected by nuclear 
activity in NGC 3079 \citep{Irwin1987}. Due to the proximity of the galaxy, NGC 3079, all the interesting features can be studied glorious detail.
\section{The Complex HI Structure of NGC 3079 }   
\begin{figure}[h]
\centering
\includegraphics[width=7cm,height=7cm,angle=-90]{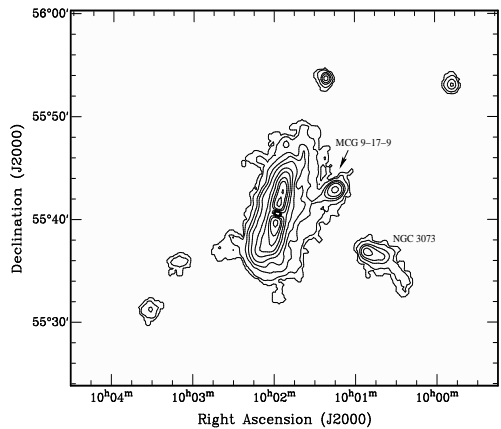}
\caption{\footnotesize Total HI distribution of NGC 3079 with HI column density contours of 0.01, 0.1, 0.3, 0.5, 1, 3, 5, 10, 30 and 50 $\times 10^{20}$ cm$^{-2}$. The HI distribution of the companion galaxies, including four new companions, is also shown here.}\label{f:totHI}
\end{figure}

Neutral hydrogen was detected in both emission and absorption. The HI emission from NGC 3079 appears to be 
more complex than thought in previous studies, with a much more extended and highly asymmetric distribution (see Fig. \ref{f:totHI}). The HI disk 
also shows an asymmetric warp in the outer regions.  A large stream 
of gas, encircling the entire galaxy was discovered, making an HI bridge between NGC 3079 and 
MCG 9-17-9, which indicates interaction between the two galaxies.  A new  companion galaxy is also discovered 
taking part in the streaming gas around the galaxy. 

\subsection{ Are MCG 9-17-9 and NGC 3073 being Ram Pressure Stripped? }   
\begin{figure}[htp]
\centering
\includegraphics[width=8cm,height=6cm]{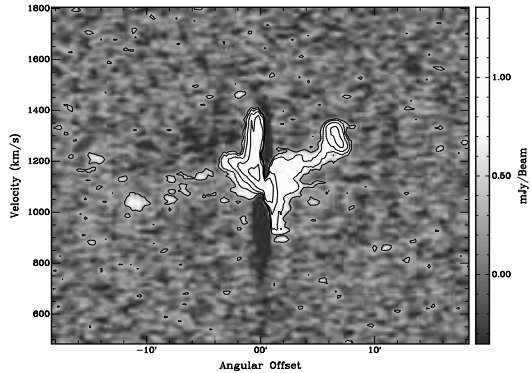}
\caption{\footnotesize Position-velocity slice taken along PA=$115^{\circ}$ showing the HI bridge between MCG 9-17-9 and NGC 3079 to the right. The ``third tail'' feature is also shown to the left. Contours are in steps of 2$\sigma$.}\label{f:interaction}
\end{figure}

MCG 9-17-9: This galaxy appears to have an interaction with NGC 3079. The HI bridge between the two galaxies is evidence for 
this (see Fig. \ref{f:interaction}), and it seems a gas tail which has been stripped off MCG 9-17-9 during the interaction. The X-ray halo of NGC 3079 is strong enough to be responsible for stripping off gas from MCG 9-17-9.\\ 
NGC 3073: Interestingly enough, an extended HI tail was seen in NGC 3073 that is much longer than seen in previous observations with significant curvature (see Fig. \ref{f:totHI}). As suggested 
by \cite{Irwin1987} the HI tail of this galaxy can be explained by ram pressure due to the supper-wind from NGC 3079. While the wind 
ram pressure seems to be responsible for the 'Cometary' appearance of NGC 3073, we have seen a drop in the surface brightness of 
NGC 3073 along the tail that can not be explained by a steady wind or constant density.  Perhaps temporal  changes in the wind 
properties are responsible for this drop. 

\subsection{A Third Tail?}    
Another interesting result (although tentative) obtained from our data is that the presence of  a third  galaxy that 
seems taking part in the interaction with NGC 3079.  An HI gas has been detected at the position of the optical galaxy SDSS 
J100311.18+553557.6 (no previous redshift was available) at a velocity of 1050 km s$^{-1}$. This HI gas exhibits a tail like structure 
forming a bridge toward NGC 3079 (see Fig \ref{f:interaction}).
\section{A fast outflow of neutral hydrogen in NGC3079? }   

Our data show strong and broad (over 600 km s$^{-1}$) HI absorption against the nuclear region
of NGC 3079. Since the the absorption is unresolved at the highest resolution of our observation, one can only tell that it originates from the gas located in front of the inner 1 kpc of the radio continuum. The HI absorption spectra show multiple-component structure with 
a mixture of broad and narrow components. The deep part of the absorption covers 
the velocity range of the observed HI emission associated with the rotating disk 
(see Fig. \ref{f:absence}).  This component has been studied in detail by \cite{Baan1995} 
and \cite{Pedlar1996}.  In addition to this already known component, we report the tentative detection of a highly blueshifted, faint component of the 
absorption that reaches velocity up to 400 km s$^{-1}$ with respect to the systemic 
velocity. This absorption is much broader than the rotation velocities observed in 
the galaxy and therefore it is likely associated with fast non-circular motions. Blueshifted absorption has to be associated to an outflow. Fast outflows of neutral 
gas have already been detected in other radio galaxies \citep[e.g;][]{Morganti2003}. 
In other radio sources where such fast cold gas has been detected the 
most likely origin for the outflow is considered to be the interaction of the jet with 
a dense interstellar medium that surrounds the radio source.  In case of NGC3079, 
however, it is more complicated to find the origin for this fast outflow. Both a 
starburst component as well as the AGN can be responsible .

\begin{figure}[ht]
\centering
\begin{minipage}[b]{0.3\linewidth}
\subfigure[]{
\includegraphics[width=7cm,height=4.5cm]{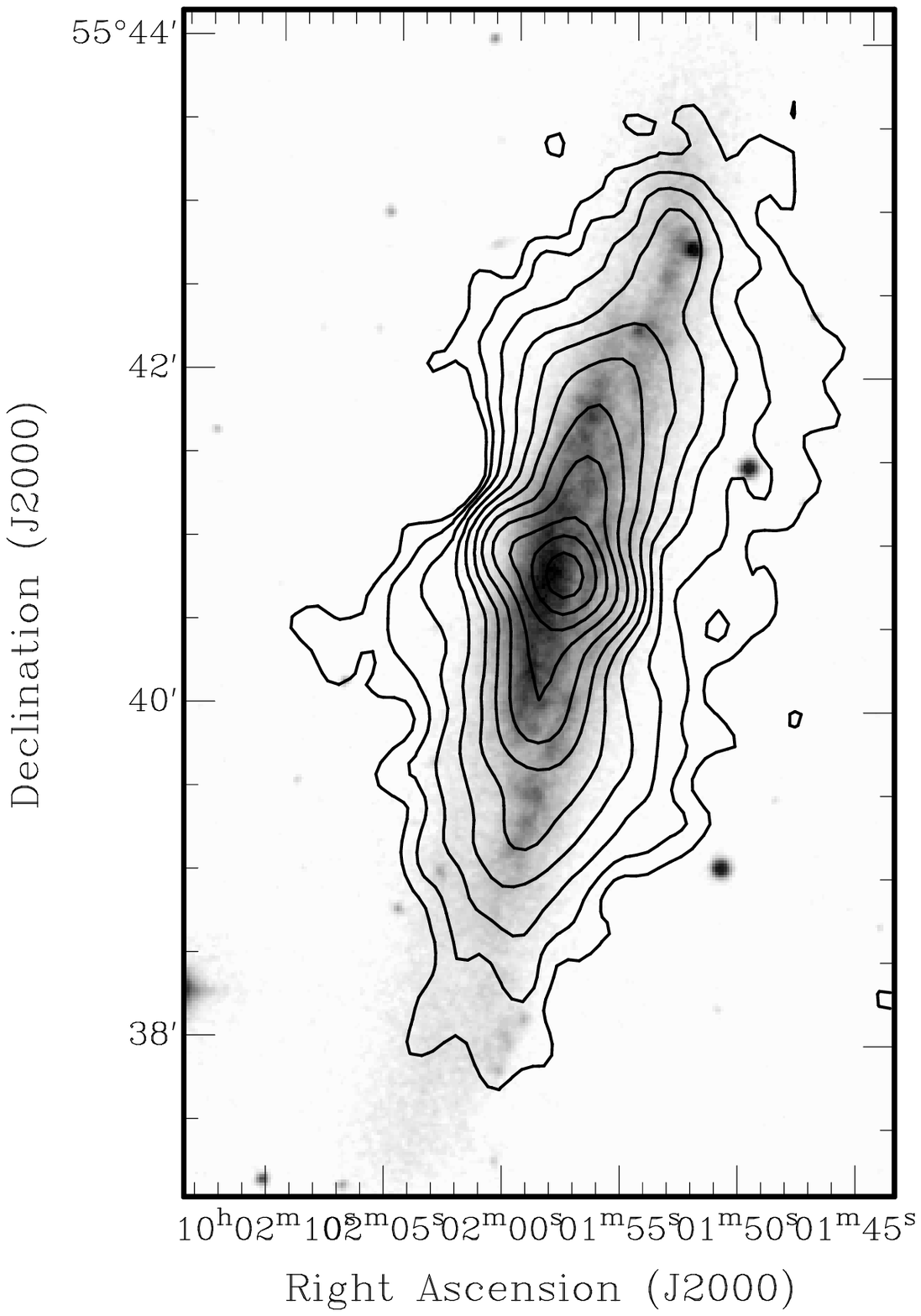}
\label{f:cont}
}
\end{minipage}
\subfigure[]{
\includegraphics[width=7cm,height=4.5cm]{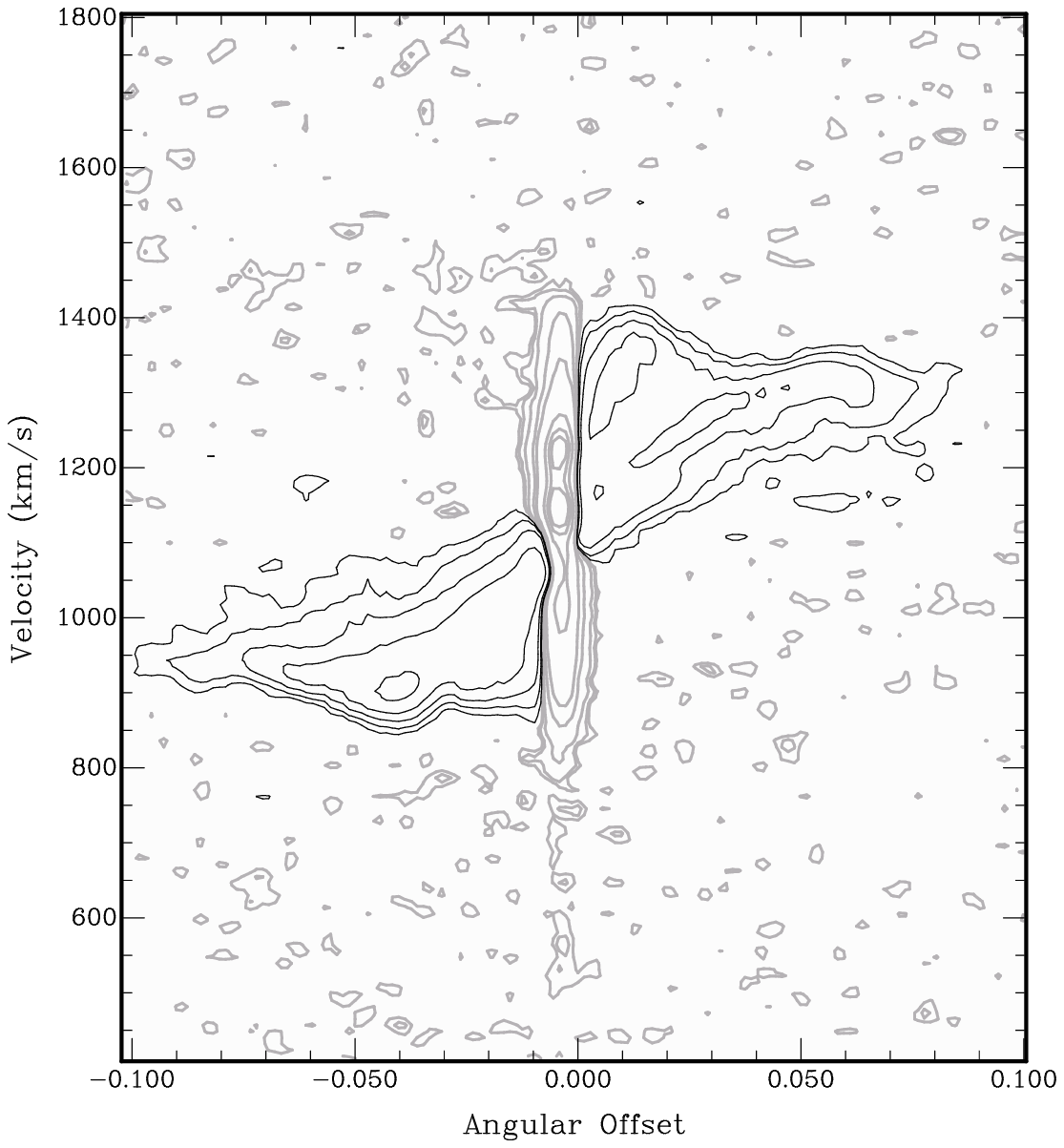}
\label{f:abspv}
}
\subfigure[]{
\includegraphics[width=4.5cm,height=4cm]{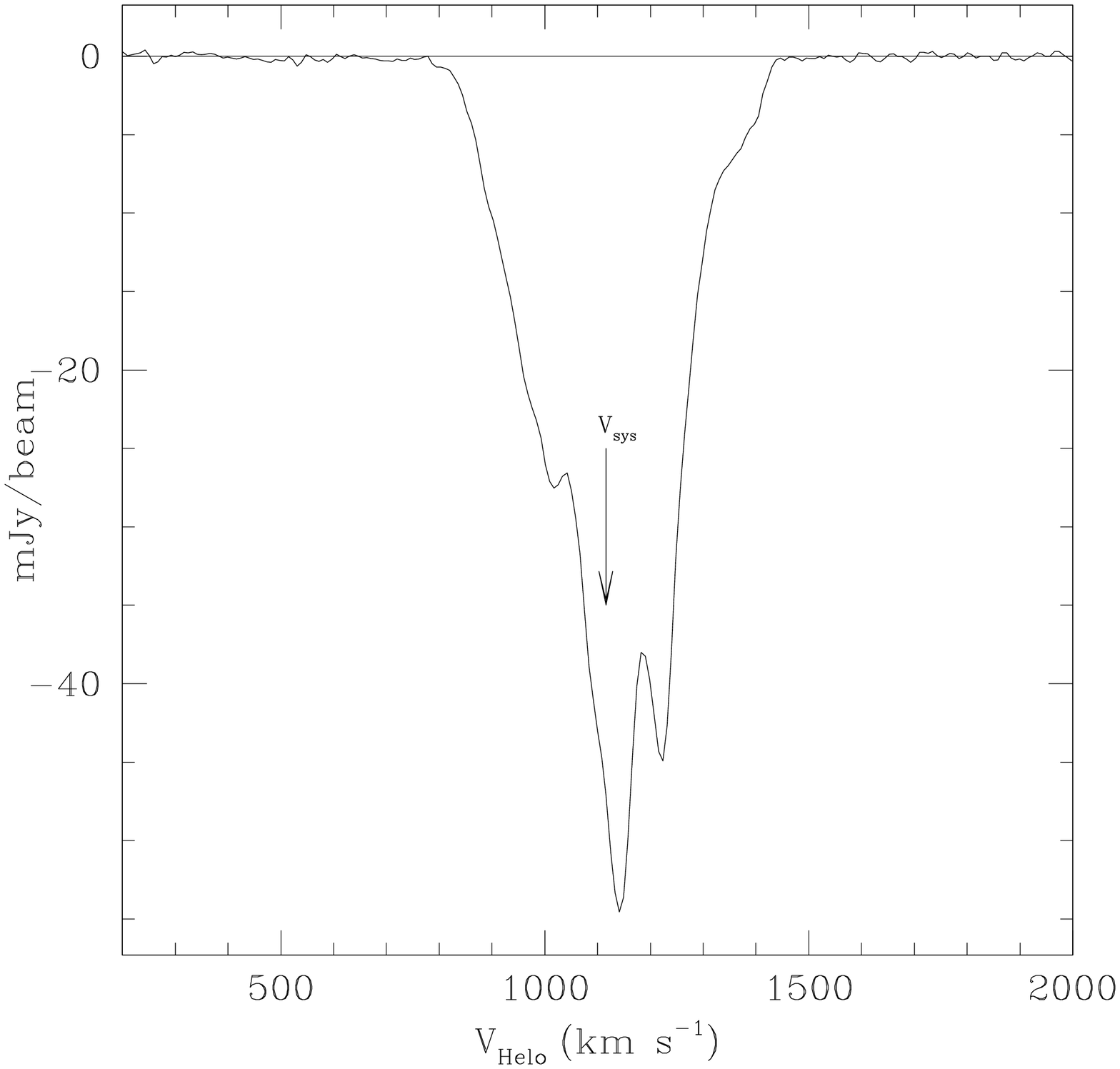}
\label{f:absspec}
}
\subfigure[]{
\includegraphics[width=3.5cm,height=3.5cm]{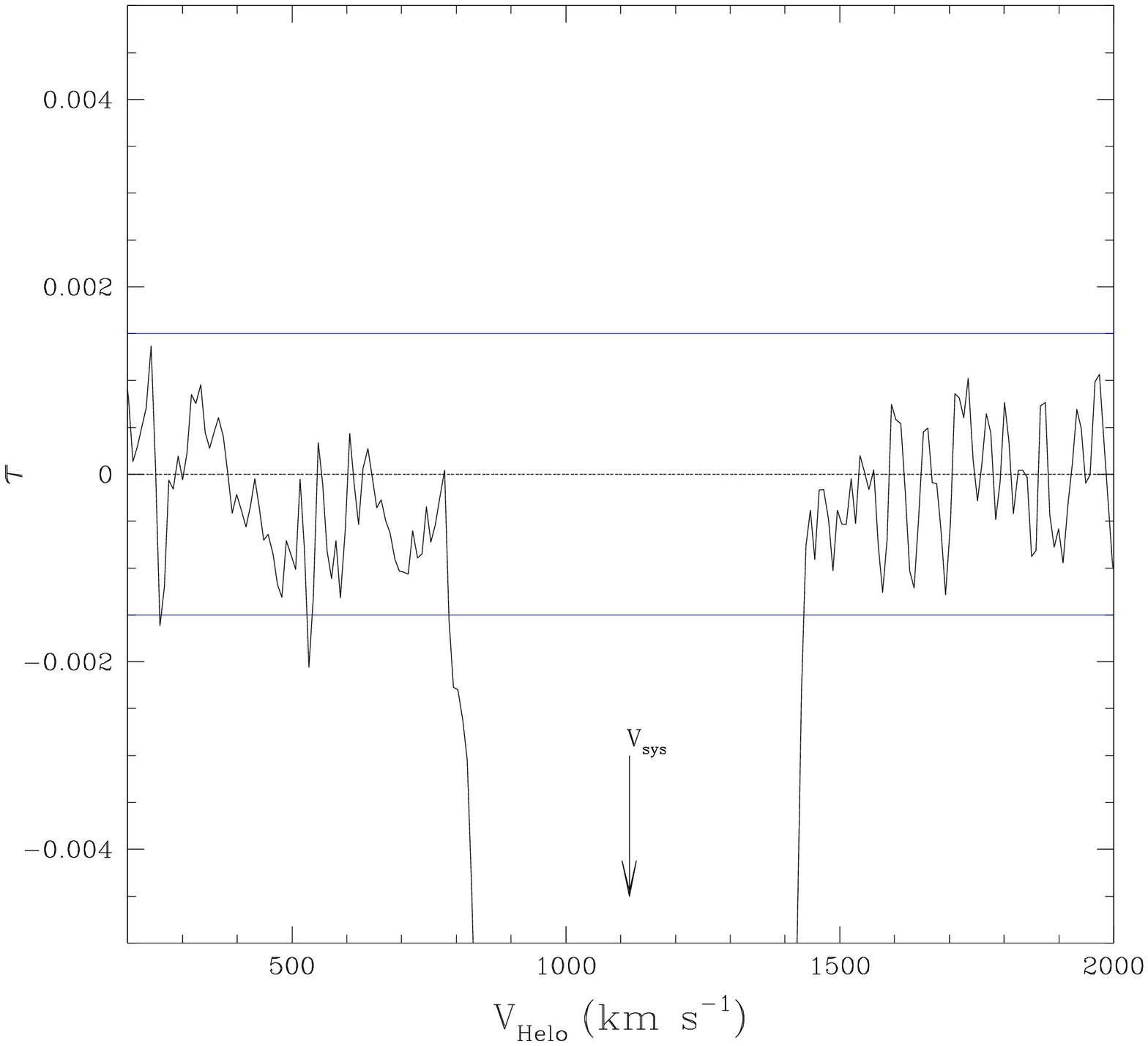}
\label{f:abszoom}
}

\caption{\footnotesize(a) Continuum map, obtained after applying  “Peeling” technique, overlaid on DSS2 image. The lowest contour level is 0.2 mJy beam $^{-1}$. Extensive continuum halo and several extension are visible. (b) Position-velocity slice taken along the major axis. Gray contours are in steps of -2 $\sigma$ level. (c) The HI absorption spectra detected against the central region of NGC 3079. (d) A zoom-in of the HI absorption optical depth.}\label{f:abs}
\end{figure}

\acknowledgements 
N. S Acknowledges financial support from the South African Square Kilometer Array (SA SKA) project and National Research Foundation (NRF).


\end{document}